\documentclass[lettersize,journal]{IEEEtran}
\usepackage{amsmath,amsfonts}
\usepackage{algorithmic}
\usepackage{algorithm}
\usepackage{array}
\usepackage[caption=false,font=normalsize,labelfont=sf,textfont=sf]{subfig}
\usepackage{textcomp}
\usepackage{stfloats}
\usepackage{url}
\usepackage{verbatim}
\usepackage{graphicx}
\usepackage{cite}
\begin{document}

\title{$Z^2$-ACT: End-to-End Verifiable Agentic Intent Control for Open 6G RAN}

\author{Sunder Ali Khowaja,~\IEEEmembership{Senior Member,~IEEE,} Kapal Dev,~\IEEEmembership{Senior Member,~IEEE,} George C. Alexandropoulos
        % <-this % stops a space
\thanks{Sunder Ali Khowaja is with School of Computer Science, Dublin City University, Dublin, Ireland. Sunderali.khowaja@dcu.ie}% <-this % stops a space
\thanks{Kapal Dev is with Department of Computer Science, Munster Technological University, Cork, Ireland. Kapal.Dev@mtu.ie}

\thanks{George C. Alexandropoulos is with Department of Informatics and Telecommunications, National and Kapodistrian University of Athens, Athens, Greece. alexandg@di.uoa.gr}}

\markboth{Journal of \LaTeX\ Class Files,~Vol.~14, No.~8, August~2021}%
{Shell \MakeLowercase{\textit{et al.}}: A Sample Article Using IEEEtran.cls for IEEE Journals}

\IEEEpubid{0000--0000/00\$00.00~\copyright~2021 IEEE}

\maketitle

\begin{abstract}
With the progression in open and disaggregated 6G radio access networks, it is expected that the system will be able to host multi-vendors. In order to host multi-vendors, it is essential that AI-assisted control loops remain safe, verifiable, and auditable under concurrent operator intents and untrusted model inputs. The existing studies address the agentic coordination, formal intent constraints, zero-trust prompt verification and cryptographic accountability in isolation, which leaves pre-realization safety, continuous semantic verification and cross-domain audit incomplete when used individually. In this regard, we propose zero-knowledge auditable control and zero-trust verifiable agentic intent architecture ($Z^2$-ACT), which integrates the aforementioned four primitives across the non-real-time and near-real-time RICs. We encode the typed Intent Contracts as operator goals while the large language model inputs are only admitted after a practical adversarial intent check. The skill sequences in the proposed study are released only when a self-management gate is satisfied while every successful commit is recorded as a binding commitment with a zero-knowledge proof. Our experimental evaluation on public ColO-RAN measurements compares the full architecture against targeted ablations and a conventional reinforcement-learning baseline. A live large language model is used in the non-real-time path to translate operator intents into Intent Contracts; we report translation accuracy, the rate of invalid or hallucinated contracts, non-real-time latency, and behavior under adversarial or misleading intents. Near-real-time control remains trace-driven on the public KPM sequences. Results indicate improved actuation filtering and attack resilience at modest latency and signaling cost inside the near-real-time envelope. Limitations of the open-loop radio replay and directions for closed-loop validation are discussed.  
\end{abstract}

\begin{IEEEkeywords}
Open RAN, 6G, Agentic AI, Intent-Based Networking, Zero-Trust, Zero-Knowledge Proofs, Trustworthiness, Near-Real-Time RIC. 
\end{IEEEkeywords}

\section{Introduction}
\IEEEPARstart{O}{ne} of the defining architectural improvements for the evolution from fifth generation (5G) communication systems to sixth generation (6G) is the shift towards open, disaggregated radio access networks (RAN). The research suggests that the Open RAN (O-RAN) has achieved greater flexibility, lower cost, and faster innovation by separating software and hardware, standardizing interfaces, and opening the control plane to multi-vendor applications, respectively \cite{ref1}. However, the openness leads to the addition of concurrent decision-makers that can significantly influence the energy policies, slice configurations, and radio resources \cite{ref2}. With the inception of large language models (LLMs) and autonomous agents, the decision-makers are being replaced by agentic controllers, which strains the control plane and increases the chances of failure modes, accordingly \cite{ref3, ref4}. Furthermore, the control plane mostly uses traditional authentication and isolation mechanisms; hence, it is not designed to handle unconstrained or hallucinated intents, prompt-injection attacks that has the capability to alter control logic, conflicting skill sequences across vendors, and the absence of cryptographically verifiable provenance, which allows administrators to audit multiple domains without seeing the raw telemetry \cite{ref5}. 

Recent research works have started addressing the aforementioned risks from different perspectives. For instance, the agentic frameworks have started treating O-RAN control entities as goal-driven agents. By doing so, the agents are allowed to operate across the non-real-time, near-real-time, and real-time layers \cite{ref6}. The frameworks also introduce structured primitives, such as multi-step planning, reusable skills, multi-horizon memory and evidence, and self-management gates that bound not only risk and uncertainty but also the budget and scope before realization. The agentic frameworks have shown measurable improvements, especially in slice life-cycle management and radio-resource control in comparison to the classical reinforcement learning baselines \cite{ref6, ref7, ref8}. Researchers have also been focusing on intent-based networking and the associated translation problem \cite{ref8}. Recently, contract-based agentic pipelines have been proposed that convert natural language operator goals into intent contracts. These contracts are schema-validated and can be audited against RAN Constraints using dual-agent loops. The audit can be performed even before the policy is dispatched, which suggests that it can eliminate harmful executions that appear when LLMs are directly realized \cite{ref9, ref10}. 

In order to cope with the security issue, researchers have proposed zero-trust prompting mechanism. The said mechanism intercepts every input that is generated by the LLM inside the near-time RIC. Each input is implicitly considered potentially adversarial; thus, it decomposes descriptive telemetry from imperative content and then decides whether to block the prompt or sanitize it. All of this process is performed while maintaining the near-real-time latency budget bound, accordingly \cite{PromptGuard}. Studies have also proposed accountability architectures that separate the real-time control path from evidence retention and trust verification. Content-addressed storage along with commitment-bound zero-knowledge proofs are used to score the decision and to trigger the policy. This combination helps in auditing multi-vendors across multiple domains without exposing model internals and compromising on performance measurements \cite{ZKTrustLLM}. O-RAN Alliance\footnote{https://www.o-ran.org/} and several other independent studies have also suggested mapping zero-trust principles onto the O-RAN architecture so that continuous verification can be performed with least privileges. The aforementioned suggestion satisfies the assumption that an adversary may already be present inside the system \cite{ref1, ZTRAN, ref11}.  

Each of the aforementioned studies addresses an important fragmentation of open multi-vendor closed-loop control control in O-RAN, however, none of the existing studies provides a single path from operator intent to audited realization. Although the agentic controllers improve explainability and long-horizon coordination, they do not bind decisions to formal pre-realization contracts or to cross-domain cryptographic provenance. On the other hand, intent contracts (IC) enforce deterministic validation of an operator goal, but they do not compose concurrent multi-vendor skills or continuously verify telemetry-derived prompts. Zero-trust prompt verification prevents the real-time decision boundary against injection, but it does not retain long-term, multi-stakeholder evidence of the actions that needs to be followed. Zero-knowledge accountability is good for trust and auditability signals, however it does not integrates with multi-timescale planning or contract governance. The gap is therefore not the absence of any mechanism rather it is the absence of an end-to-end verifiable intent-to-realization chain, in which the typed intent is admitted only after semantic checks are performed, it is then realized only through gated skill sequences under concurrent multi-vendor constraints, and then recorded so that any authorized part can later verify the decision without exposing raw telemetry. The proposed study develops that end-to-end chain in the form of $Z^2$-ACT. We perform a comparative analysis of the existing works, especially in the domain of agentic O-RAN, intent-contract, zero-trust and zero-knowledge approaches along the properties required for end-to-end chain in Table 1, accordingly. 

% \usepackage{graphicx}
\begin{table}[]
\centering
\caption{Comparison with the existing approaches ($\checkmark$ = addressed; $\Delta$ = partial; --- = not addressed). It should be noted that the "Partial" $\Delta$ indicates that the cited work discusses the issue or provides a related mechanism, but does not close the property into aa single intent-to-realization pipeline. $Z^2$-ACT claim is the chain, not the invention of each block in isolation. }
\label{tab:my-table}
\resizebox{\columnwidth}{!}{%
\begin{tabular}{|l|l|l|l|l|l|}
\hline
\multicolumn{1}{|c|}{\textbf{Property}} & \multicolumn{1}{c|}{\textbf{\cite{ref6}}} & \multicolumn{1}{c|}{\textbf{\cite{ref9}}} & \multicolumn{1}{c|}{\textbf{\cite{PromptGuard}}} & \multicolumn{1}{c|}{\textbf{\cite{ZKTrustLLM}}} & \multicolumn{1}{c|}{\textbf{$Z^2$-ACT}} \\ \hline
\textbf{\begin{tabular}[c]{@{}l@{}}Formal pre-actuation\\ Intent Contract\end{tabular}} & --- & $\checkmark$ & --- & --- & $\checkmark$ \\ \hline
\textbf{\begin{tabular}[c]{@{}l@{}}Continuous zero-trust\\ check on LLM-bound\\ inputs\end{tabular}} & --- & --- & $\checkmark$ & --- & $\checkmark$ \\ \hline
\textbf{\begin{tabular}[c]{@{}l@{}}Gated skill sequencing\\ / self-management \\ under Near-RT constraints\end{tabular}} & $\checkmark$ & --- & --- & --- & $\checkmark$ \\ \hline
\textbf{\begin{tabular}[c]{@{}l@{}}Multi-vendor conflict\\ handling (priority + scope)\end{tabular}} & $\Delta$ & $\Delta$ & --- & --- & $\checkmark$ \\ \hline
\textbf{\begin{tabular}[c]{@{}l@{}}Post-realization ZK \\ commitment + proof (no\\ raw telemetry exposure)\end{tabular}} & --- & --- & --- & $\checkmark$ & $\checkmark$ \\ \hline
\textbf{\begin{tabular}[c]{@{}l@{}}End-to-end intent $\rightarrow$\\ verified realization $\rightarrow$\\ audit chain\end{tabular}} & --- & --- & --- & --- & $\checkmark$ \\ \hline
\textbf{\begin{tabular}[c]{@{}l@{}}Evaluation on public O-RAN \\ traces\end{tabular}} & $\Delta$ & $\Delta$ & $\Delta$ & $\Delta$ & $\checkmark$ \\ \hline
\end{tabular}%
}
\end{table}

\begin{figure*}[!t]
\centering
\includegraphics[width=0.8\linewidth]{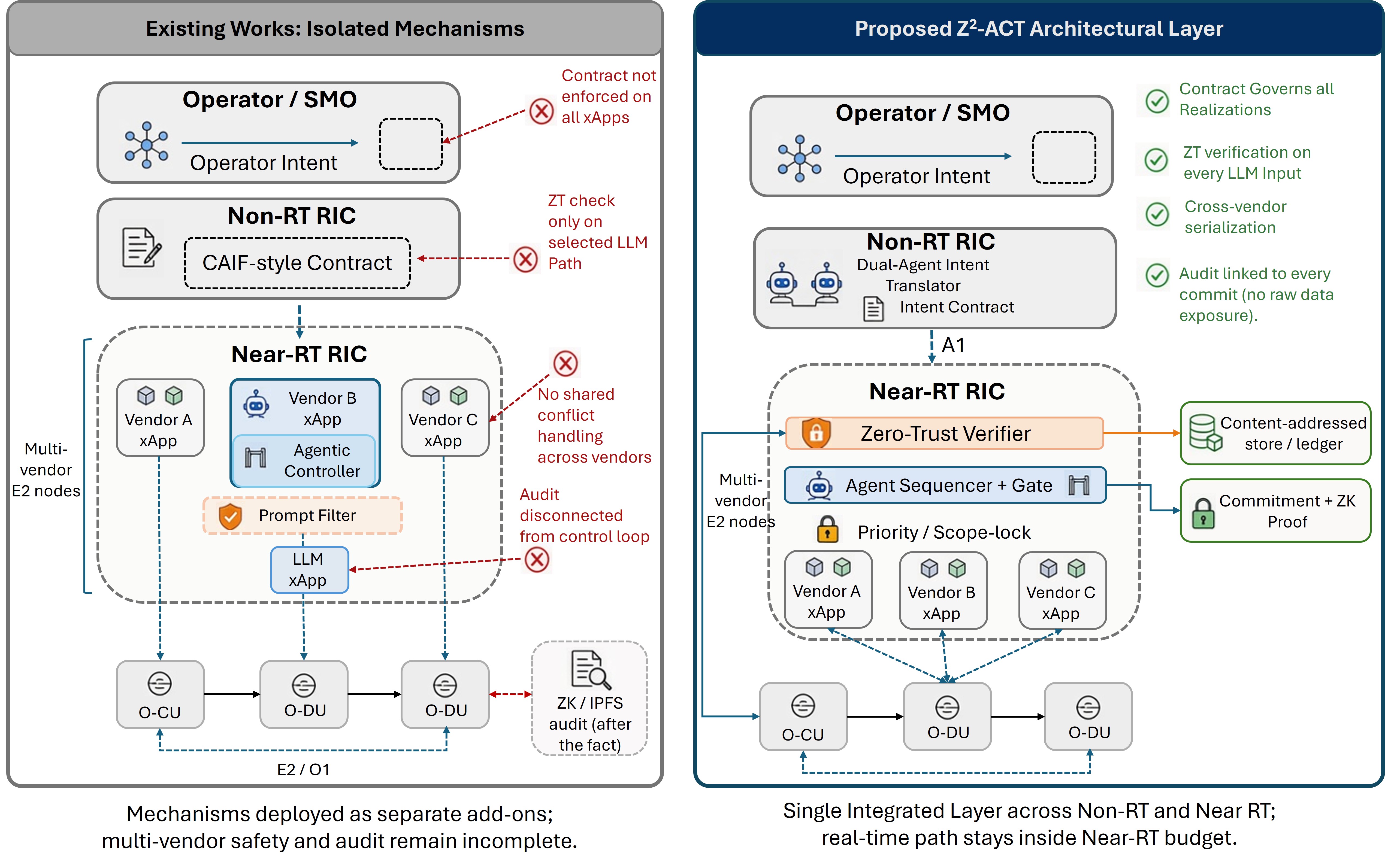}
\caption{Real-world O-RAN Deployment View: Existing approaches vs $Z^2$-ACT}
\label{fig_1}
\end{figure*}

Based on the aforementioned observations, the proposed study raises three concerns. The first is related to the four primitives, i.e. agentic multi-timescale control, formal intent contracts, zero-trust verification of LLM-bound inputs, and commitment-bound zero-knowledge audit. The proposed study explores whether the four primitives can be composed into a single architectural layer that not only guarantees pre-realization safety but also continuous adversarial-intent checking, multi-vendor conflict handling, and cross-domain auditability. The second is to explore whether the formal properties, such as constraint satisfaction of contracts, bounds on adversarial intent probability, soundness of the commitment scheme, and latency overhead under real-time constraints, can be used for the aforementioned composition of four primitives. Lastly, the third is to explore whether the resulting architecture can improve service-level agreement satisfaction, attack resilience, and audit verification in comparison to the conventional baselines, especially when evaluated on publicly available multi-cell O-RAN traces. 

In this regard, we proposed $Z^2$-ACT, a Zero-Knowledge Auditable Control and Zero-Trust Verifiable Agentic Intent Architecture that focuses on the intent and control management layer of the Open-RAN architecture in a unified manner. The proposed $Z^2$-ACT introduces a verifiable agentic intent (VAI) that embeds contract-governed intent translation, zero-trust verification, agentic skill sequencing, and commitment-bound zero-knowledge audit in a unified architecture. The proposed architecture will be interoperable with existing A1, E2, and O1 interfaces while adding an evidence bus that supports multi-vector provenance without exposing raw telemetry. The issues concerning existing works and how the proposed work addresses them is illustrated in Figure 1.  

The remainder of the paper is structured as follows. Section II and Section III review the fundamentals of agentic control in O-RAN along with the verification and accountability primitives. Section IV presents the proposed $Z^2$-ACT architecture, its analytical model, security analysis and the composition policies. Section V presents the evaluation setup based on ColO-RAN and OpenRAN Gym datasets. Section VI presents the experimental results, and lastly Section VII concludes the study. 

\section{Fundamentals of Agentic Control in O-RAN}
In O-RAN, agentic control assumes the control plane as goal-driven agents rather than reactive and isolated machine learning models. An agentic controller in O-RAN starts with a goal, decomposes the goal into sequences of actions, observes the outcomes, and then updates the parameters and states over multiple time scales, in contrast to the conventional xAPPs that map a fixed state vector into a set of discrete actions \cite{ref4}. This paradigm shift occurred after the studies \cite{ref6, ref7} suggested that the O-RAN with multiple vendors and multiple objectives require long-horizon coordination, which the classical reinforcement learning loop struggles with under the timing and observability constraints. 

Generally, the building blocks of an agentic O-RAN controller can be characterized into four primitives. The first is the plan-act-observe-reflect cycle. This primitive varies for non-real-time and near-real-time RICs. For instance, an agent in a non-real-time RIC interprets an operator intent, consults long-term memory and then produces a structured plan, while the same agent in a near-real-time RIC executes shorter skill sequences, observes key performance measurements, and reflects on whether the plan remains valid. The second refers to the skills as tool use. The skills can be referred to as bounded, reversible realizations that map onto standardized O-RAN interfaces like A1 policy updates, E2 control messages, or O1 configuration changes. Concurrent agents using skills can detect and order conflicts without centralized arbitration, as each skill carries an a priori declaration of its expected effect and cost, accordingly \cite{ref6, ref1}. The third primitive corresponds to the multi-horizon memory and evidence. For instance, short-term state supports near-real-time gating, whereas long-term knowledge retains validated skill compositions, and episodic records store decision-outcome pairs, respectively. Evidence, which can be described as a compact record linking the goal, the selection action, and the observed outcome, is generated at every commit. This evidence is an important characteristic of an agentic controller as it supports post-audit and regulatory reporting without compromising raw subscriber data. The last primitive refers to the self-management gate, suggesting that the agent evaluates the predicted risk, uncertainty, resource budget, spatial scope, and commit rate before even any realization takes place. Once the agent thinks that all guards will pass, then and only then it proceeds with the selected action; until then, it shrinks the step, delays it and rolls it back. Therefore, we can say that the gate operationalizes surrogates for stability under the latency budgets of the near-real-time RIC, which is effective in practical scenarios \cite{ref6, ref4}.

The aforementioned primitives can be naturally mapped onto the O-RAN control hierarchy. The reason for this claim is that the LLM reasoning is confined to the non-real-time layer, which is susceptible to inference latency. This separation is preserved in $Z^2$-ACT as in the proposed study, the near-real-time zero-trust verifier does not host the LLM rather it only admits or blocks LM-bound inputs before skill sequencing, accordingly. The agents that operate at the near-real-time level operate without large-model calls, relying on lightweight skill sequencing and gating. This helps to keep the control loop intact within a 10 ms - 1 sec time bound. Some agents, like dApp-level agents, which operate in real time, are restricted to even narrower and pre-validated skill sets. Such decomposition not only preserves the timing and interface contracts of the O-RAN Alliance specifications but also adds an organizational layer that can coordinate multiple underlying optimizers \cite{ref7, ref12}. 

Even with all the advantages mentioned above, agentic control expresses several gaps. For instance, the plans generated by LLMs might violate hard resource ceilings or service-level agreements unless an external validation step is carried out. Another example corresponds to concurrent multi-vendor agents, as they can still produce conflicting realizations if the only conflict resolution mechanism is local gating. Furthermore, the telemetry in the agent's context window is vulnerable to prompt injection attacks. Lastly, the evidence can help in performing local audit but fails to provide cryptographically verifiable provenance that can be checked by independent administrative domains without exposing the underlying measurements. We explore the mechanisms that address these limitations in the subsequent section. 

\section{Fundamentals of Intent Contracts, Zero-trust Prompting, and Zero-Knowledge Accountability}
Although the organizational structure for multi-timescale control is provided by the agentic primitives, three additional mechanisms help in addressing the safety, verification, and auditability issues associated with multi-vendor settings. These mechanisms are reviewed in the subsequent subsections, accordingly. 
\subsection{Intent Contracts}
An intent contract can be defined as a formal, machine-checkable representation of an operator goal coupled with the constraints that need to be satisfied for an admissible realization. Unlike the natural-language instructions that are provided in free form, the intent contract takes into account the information regarding temporal scope, resource ceilings, throughput or latency bounds, target slice identifiers, and permissible primitives. Generally, the contract employs a dual-agent pipeline, i.e., a profiling agent and an evaluator agent. The former agent extracts candidate parameters from the operator's input and the current network state, while the latter agent verifies the feasibility of the resulting specification with respect to the live RAN configuration, ensuring it does not violate the safety or isolation rules. The contract for the downstream policy generation is released only if both agents come to a consensus agreement \cite{ref9, ref10}. 

The primary reason for employing the intent contracts is that it decouples the probabilistic nature of LLM interpretation from the non-contextual enforcement of RAN constraints. Upon the acceptance of the contract, the agents are allowed to optimize, but only within the declared feasible set. The main limitation of the intent contract is that it validates a single intent in isolation; it does not detect or resolve conflicts concerning the request of overlapping resources by itself, which might arise from multiple concurrent contracts originating from different vendors.  
\subsection{Zero-Trust Prompting}
The primary assumption associated with zero-trust prompting in $Z^2$-ACT is that every LLM-bound input that can influence the near-real-time skill sequencer is considered to be potentially adversarial. In the proposed $Z^2$-ACT, the LLM reasoning that is used to translate operator intent into IC is confined to the non-real-time RIC, therefore, no large LLM is required to be executed inside the near-real-time control loop. In this regard, we assume that the LLM bound inputs include (i) high-level plans and A1 policies that can be produced with the assistance of LLM in the non-real-time layer, and (ii) prompt-style summaries or telemetry driven natural-language which can affect near-real-time decision logic. The rationale is that the telemetry aggregates, shared state variables, and even previously generated policies can be engineered or poisoned to compel the model to produce unsafe control decisions. The zero-trust verification component, in this case, acts as the first line of defense. It intercepts the prompt before it reaches the model; it decomposes the prompt into imperative content that represents requested actions and descriptive content corresponding to measurements and topologies. After the decomposition, it computes an adversarial-intent score. Only prompts that do not pass the calibrated threshold can be forwarded; otherwise, the prompts are either blocked or sanitized, accordingly \cite{PromptGuard}. 

The constraint concerning zero-trust prompting is that the verification is performed inside the near-real-time RIC; thus, it needs to be performed within the control budget. In this regard, the implementation of zero-trust prompting is realized as a lightweight xApp, hence avoiding heavy model calls. Existing studies have evaluated zero-trust prompting on open testbeds and have shown that it can achieve high detection accuracy while complying with the latency requirements of the O-RAN loop. The only limitation of zero-trust prompting is that it only protects the input boundary of the language model. Therefore, it does not generate long-term, multi-stakeholder evidence of the decisions that are executed. 
\subsection{Zero-Knowledge Accountability}
The main problem that zero-knowledge accountability addresses is the cross-domain audit without exposing the raw data. After the control decision, a compact evidence digest is formed along with a commitment to the decision that was made. In order to attest to the satisfaction of a declared policy predicate and the correctness of the commitment, a zero-knowledge proof is generated. This zero-knowledge proof does not reveal any underlying information, including key performance measurements or model internals. The proofs and commitments are then stored in a content-addressable repository and might be recorded on a permissioned ledger (depending on SLAs or preferences) so that the claims can be verified at later stages, accordingly \cite{ZKTrustLLM}. It should be noted that the architecture employing zero-trust accountability has to separate the real-time control path from the evidence and trust planes deliberately so that the cryptographic operations do not affect the near-real-time latency. The main limitation of zero-knowledge accountability is that it focuses only on post-facto auditability rather than on preventing an unsafe realization in the first place. 

Considering the aforementioned three mechanisms, they do provide pre-realization constraint enforcement, continuous semantic verification of model inputs, and post-realization cryptographic provenance, but neither do they replace the organizational structure of agentic control nor do they automatically coordinate with one another. In this regard, we propose the $Z^2$-ACT architecture that explores how the four primitives can be integrated into a unified and coherent architecture. 

\begin{figure*}[!t]
\centering
\includegraphics[width=\linewidth]{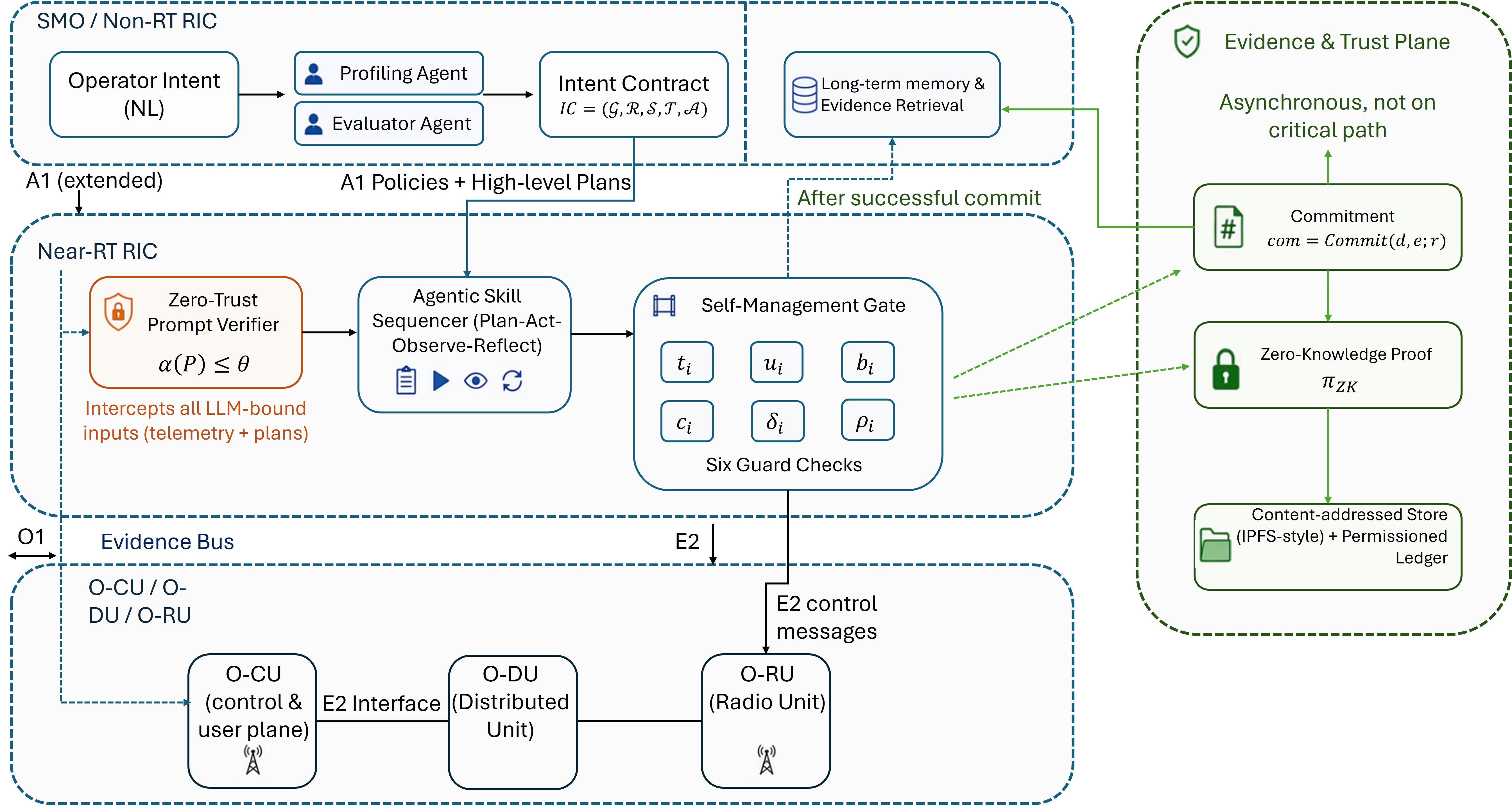}
\caption{Proposed $Z^2$-ACT Architectural Layer}
\label{fig_2}
\end{figure*}

\section{$Z^2$-ACT architecture}
We previously reviewed the four primitives of open multi-vendor control in Sections II and III, that explains a distinct aspects of open multi-vendor control, including organizational structure and multi-timescale coordination (agentic primitives), pre-realization constraint enforcement (intent contracts), continuous semantic verification of model inputs (zero-trust prompting), and post-realization cryptographic provenance (zero-knowledge accountability). Existing works have explored the aforementioned primitives, mostly in isolation, therefore, the question that remains is whether they can be integrated into a single architectural layer that simultaneously guarantees safety, continuous adversarial-intent checking, multi-vendor conflict handling, and cross-domain auditability, while being compatible with the timing and interface contracts of the O-RAN Alliance specifications. 

The proposed architecture is illustrated in Figure 2. The proposed single architectural layer is intended to be embedded at the non-real-time RIC / SMO layer, where a natural-language operator intent is processed by a dual-agent pipeline (profiling agent and evaluator agent), hence producing a formal Intent Contract $IC = (\mathcal{G}, \mathcal{R}, \mathcal{S}, \mathcal{T}, \mathcal{A})$. Here, the $\mathcal{G}$ is referred to as a set of goal predicates such as minimum throughput or maximum latency, the $\mathcal{R}$ corresponds to the set of hard resource ceilings, the $\mathcal{S}$ refers to the spatial scope like cells or PRB groups, the $\mathcal{T}$ refers to the temporal validity interval, and $\mathcal{A}$ corresponds to the set of permitted skill identifiers. This implies that only the contracts that satisfy the typed feasibility constraints are released. In the same layer, the long-term memory mechanism and the evidence retrieval are also integrated. The resulting A1 policies and high-level skill plans are handed over to the near-real-time RIC, accordingly. The LLM reasoning will remain confined to the non-real-time layer, therefore, the near-real-time path will not host the model. In contrast, the near-real-time RIC hosts a zero-trust prompt verifier that intercepts every LLM-bound input, i.e. A1 plans produced with LLM assistance and telemetry-driven prompt-style summaries, whcih could influence the realization. The verifier then computes an adversarial intent score $\alpha(P)$ and passes it only if $\alpha(P) \leq \theta$. The notation $\theta \in$ (0,1) is a calibrated threshold. Once the inputs are passed, it reaches the agentic skill sequencer that operates under a self-management gate. Before any skill is released, as an E2 control message towards the O-CU/O-DU, the gate predicts the risk $t_{i}$, uncertainty $u_{i}$, budget usage $b_{i}$, explanation consistency $ec_{i}$, spatial scope $\delta_{i}$, and recent commit rate $\rho_{i}$, accordingly. Once the commit is successful, an asynchronous Evidence and Trust Plane forms a cryptographic commitment to the decision and evidence digest while generating a zero-knowledge proof of consistency with the accepted contract and the gate conditions. Both of the objects are then stored in a content-addressable repository (optionally they can be anchored on a permissioned ledger as well). During the whole process, the real-time control path never gets delayed by the cryptographic operations, but at the same time, any authorized multi-vendor stakeholder can verify the proof without the raw key-performance measurements being exposed. The proposed architecture remains interoperable with the existing AI, E2, and O1 interfaces with only one addition, i.e., a lightweight evidence bus.  
\subsection{System Model and Formulation}
In the proposed work, we fix the discrete decision epochs of the near-real-time RIC and index them as $i \in \mathbb{N}$. The controller observes a network state vector denoted as $s_{i} \in \mathcal{S} \subset \mathbb{R}^{n_{s}}$ at each epoch. The components of the network state vector include key performance measurements such as throughput, latency, physical-resource-block utilization and related quantities, which arrive over the E2 interface. An operator then provides a high-level goal in the form of a natural-language string denoted by $I$. The dual agent pipeline that resides in the non-real-time RIC takes the input $(I, s_{i})$ and tries to produce a machine-checkable IC. From this point onwards, all the subsequent layers will act upon the contract object only. Therefore, the contract serves as a formal bridge between the operator's informal request and the constrained optimization problem that the agents in near-real-time RIC will try to solve. In this study, we formally define the Intent Contract as $IC = (\mathcal{G}, \mathcal{R}, \mathcal{S}, \mathcal{T}, \mathcal{A})$, where $\mathcal{G} = \{g_{k}: \mathcal{S} \rightarrow \mathbb{R}\}_{k=1}^{K}$ represent the finite collection of goal predicates like lower bound on slice throughput, $\mathcal{R} = \{r_{l}: \mathcal{S} \rightarrow \mathbb{R}\}_{l=1}^{L}$ corresponds to a finite collection of hard resource ceilings, $\mathcal{S} \in \{1,...,N_{cell}\}$ refers to the spatial scope expressed as a set of cells of resource block groups, $\mathcal{T} = [t_{start}, t_{end}]$ corresponds to the temporal interval during which the contract remains valid, and lastly $\mathcal{A}$ refers to the catalogue of skill identifiers that the contract authorizes. If none of the tuples satisfies the live RAN configuration, the dual agent pipeline returns a rejection symbol, which suggests that no further realization is attempted, accordingly. 

The decision at epoch $i$, represented as $d_{i}$, is a finite ordered sequence of skills. An individual skill, denoted by $\sigma \in \mathcal{A}$ is a map such that $\sigma: \mathcal{S} \times \Theta_{\sigma} \rightarrow \mathcal{S}$ transforms the current state based on the parameter set $\Theta_{\sigma}$. The decision will only be processed for the contract whose state $s'$ when executing $d_{i}$ satisfies every goal predicate, respects every resource ceiling, lies inside the declared spatial scope, and occurs inside the declared temporal interval as shown in equation 1.  
\begin{equation}
    d_{i} \models IC \iff
    \begin{cases}
        g_{k}(s') \geq 0 & \forall g_{k} \in \mathcal{G},\\
        r_{l}(s') \leq 0 & \forall r_{l} \in \mathcal{R}, \\
        supp(d_{i} \subseteq \mathcal{S}), \\
        time(d_{i}) \subseteq \mathcal{T}
    \end{cases}
\end{equation}
Therefore, admissibility is purely a declarative property of the pair $(d_{i}, IC)$, however, it should be noted that it does not yet incorporate timing, uncertainty or concurrent load. 

The proposed method first makes the input from produced from LLM that can influence the near-real-time skill sequencer undergo a zero-trust check before it is examined for the admissibility condition. It should be noted that the near-real-time verifier does not execute the model rather it protects the admission of text-form or prompt-style objects so that only the inputs with $\alpha(P) \leq \theta$ can affect the release predicate. Each input, denoted by $P$ is decomposed into a descriptive component $P_{desc}$ and an imperative component $P_{impr}$. The former refers to the telemetry and topology, while the latter corresponds to the requested actions. A detector then returns a scalar score, i.e. $\alpha(P) \in$ [0,1]. It should be noted that the score is an idealized modelling abstraction of the detection mechanisms, which are reported in the literature. The study does not claim this to be a rigorously calibrated conditional probability. The input undergoes examination for admissibility condition only when $\alpha(P) \leq \theta$
for a configurable threshold $\theta \in (0,1)$. This makes sure that every plan that reaches the skill sequencer has cleared the zero-trust check, accordingly. 

It is assumed that even an admissible decision can be unsafe under the current load or uncertainty; therefore, in the proposed work, the self-management gate evaluates six scalar quantities that capture the predicted risk, uncertainty, resource consumption, explanation consistency, spatial extent, and recent realization rate, as shown in equation 2. 
\begin{equation}
\begin{aligned}
    & t_{i} = t(s_{i}, d_{i}), u_{i} = u(s_{i}, d_{i}),\\
    & b_{i} = b(s_{i}, d_{i}), ec_{i} = ec(s_{i}, d_{i}),\\
    & \delta_{i} = |supp(d_{i})|,  \rho_{i} = \frac{1}{W} \sum_{j=i-W+1}^{i} 1_{\{d_{j} released\}} \\
\end{aligned}
\end{equation}
The aforementioned six numbers define a feasible region inside the set $\mathcal{D}$ of all finite skill sequences. The feasible region can formally be defined as follows:
\begin{equation}
    \begin{aligned}
        \mathcal{F}(\Theta) = & \{d \in \mathcal{D}: t(d) \leq \alpha, u(d) \leq \beta, b(d) \leq \gamma, \\
        & c(d) \geq \tau, \delta(d) \leq S, \rho(d) \leq R\},
    \end{aligned}
\end{equation}
where $\Theta = (\theta, \alpha, \beta, \gamma, \tau, S, R)$ is the vector of thresholds that are active for the current epoch. The gate admits a decision only if it lies inside the region. Together, the aforementioned three checks yield the release predicate that governs the realization, which is formally defined in equation 4.
\begin{equation}
    \begin{aligned}
        Release(d_{i}, IC) \equiv & (d_{i} \models IC) \land (d_{i} \in \mathcal{F(\Theta)}) \\
        & \land (\forall P \text{ that influenced } d_{i}: \alpha(P) \leq \theta)
    \end{aligned}
\end{equation}
The corresponding E2 control message is then dispatched to the O-CU or O-DU only when the evaluation of the predicate returns "True". The proposed architecture ensures that architecture should produce an auditable record after a successful release, while not exposing any raw measurements. In this regard, the Evidence and Trust Plane should produce an auditable record that does not expose raw measurements. We instantiate the plane as $com_{i} = Commit(d_{i}, e_{i};r_{i}) = g^{d_{i}}h^{e_{i}}u^{r_{i}}$, which is also referred to as Pedersen commitment \cite{pedersen}. In the aforementioned expression, the $g$ refers to a generator of the prime-order group $\mathbb{G}$ (public parameter), $h$ refers to a second generator of $\mathbb{G}$, which is chosen so that $log_{g}h$ is unknown, and the notation $u$ refers to a third generator of $\mathbb{G}$, with unknown discrete log relative to $g$. The $r_{i}$ in the aforementioned expression is sampled uniformly. Under the discrete-logarithm assumption, the commitment is binding, i.e. a probabilistic polynomial-time adversary that cannot open $com_{i}$ to two distinct pairs $(d_{i}, e_{i})$ and hiding the distribution of $com_{i}$ while being independent of $d_{i}$ and $e_{i}$ when $r_{i}$ is uniform. In the proposed work, the proof system is a Groth16 zk-SNARK \cite{zksnark} with a one-time setup that comprises of a proving key and a public verification key $vk$. The public inputs are $(com_{i}, H(IC),i$ while the private witness $(d_{i}, e_{i}, r_{i})$ are paired with the data that is needed to show that $Release(d_{i},IC) = True$. The assumption is that the $com_{i}$ will only open to $(d_{i},e_{i})$ and that the release predicate holds for the accepted IC and the gate thresholds at epoch $i$. Formally, the verification process is defined in equation 5.  
\begin{equation}
    Verify(com_{i}, \pi_{ZK}^{(i)}, vk) = 1,
\end{equation}
The aforementioned equation suggests that the verification succeeds if and only if the values are consistent with the commitment while satisfying the release predicate. The zero-knowledge of the SNARK ensures that a verifying proof never reveals the decision as well as the evidence digest. It should be noted that we do not re-prove the cryptographic properties of Pedersen commitments or Groth16 rather we inherit them from the standard constructions \cite{ref13, ref14} and use a 128-bit security parameterization on a pairing-friendly curve. It should also be noted that the proof generation is asynchronous and does not appear in $L_{i}$.

Subsequently, the near-real-time path is subject to a hard latency constraint. The total latency of epoch $i$ is computed as the sum of the verifier latency, the gate-evaluation latency and the E2 dispatch latency, which are the wall-clock components as shown in equation 6.
\begin{equation}
    L_{i} = L_{ver}(P) + L_{gate}(d_{i}) + L_{E2}
\end{equation}
where the $L_{ver}(P)$ refers to the clock time required by the zero-trust prompt verifier to decompose the input $P$ and to evaluate the score $\alpha(P)$, the $L_{gate}(d_{i})$ corresponds to the clock time required to compute the six scalar gate quantities and to evaluate the release predicate $Release(d_{i}, IC)$, and the $L_{E2}$ refers to the clock time required to encode the authorized skill sequence into an E2 control message and to dispatch it towards the O-CU and O-DU. The aforementioned latencies are obtained using direct measurement on the target near-real-time RIC platform. It should be noted that they are not derived from complexity expressions. The proposed architecture is required to keep $L_{i} \leq L_{max}$ for every epoch, where $L_{max}$ lies inside the interval [10 ms, 1 s], which is prescribed by the O-RAN specification. We do not consider the cryptographic work of the Evidence and Trust Plane in $L_{i}$ as it is performed in an asynchronous manner. 

\subsection{Invariants and Derived Properties}
We have explained how the three independent filters, i.e., contract admissibility, zero-trust admission and gate feasibility, are encoded. In this subsection, we explain the invariants that follow directly once the external assumptions (soundness of the dual-agent pipeline, adequate performance of the practical score $\alpha$, correct enforcement of the gate inequalities, and the cryptographic properties of the commitment and proof system) are granted. 

The first invariant suggests that the release predicate is both necessary and sufficient for realization, i.e., a skill sequence reaches the E2 interface only if $Release(d_{i}, IC)$ returns "True". However, if the predicate violates the IC, fails the zero-trust test, or lies outside the gate region, it will never be applied to the radio access network. The second invariant is related to auditability. It is assumed that if the verifier accepts the proof $\pi_{ZK}^{(i)}$ for the commitment $com_{i}$, then there should exist a decision-evidence pair which satisfies the release predicate at the moment of release. As per the zero-knowledge property, verification runs in time polynomial in the size of the proof and does not reveal the decision or the evidence digest, respectively. Therefore, the cross-domain stakeholders can check the correctness without recovering any raw measurements.  

In practical scenarios, the spatial scopes might overlap when multiple contracts are active at the same time. We use a simple priority rule in order to resolve such conflicts, which is that safety and compliance are always handled first, followed by service-level agreement recovery contracts, and finally by efficiency-oriented contracts. We employ scope lock in the case where two contracts affect the same cells or resource blocks. The scope lock ensures that the lower-priority contract is considered only after the high-priority one has been either accepted or rejected. Each contract is checked against its own IC and against the current gate conditions. Hence, the safety property is preserved for every contract while the release decision remains well-defined. 

\subsection{Provisioning policies}
In the proposed work, the safety and auditability properties depend on how the numerical thresholds are set and how the contracts are handled when submitted in a concurrent manner. In this regard, we define a small set of provisioning rules. The threshold vector $\Theta$ is adjusted at runtime, suggesting that when the uncertainty score or the resource budget usage remains above a hysteresis band for a dwell time $W_{d}$, every component of $\Theta$ is tightened by a fixed step. Conversely, every component of $\Theta$ is relaxed by the same step when the same quantities remain inside the normal operating band for the same dwell time. Therefore, the dwell time prevents the thresholds from oscillating.

A decision is written into long-term memory only if the Euclidean distance between the observed and predicted key-performance measurements falls below a tolerance denoted by $\varepsilon_{KP}$. If the distance goes beyond $\varepsilon_{KP}$, the decisions are discarded automatically. In the proposed work, the query cost grows only logarithmically with the number of stored entries and stays negligible compared with the non-real-time control interval because the retrieval uses nearest-neighbour search over a fixed-dimensional embedding of the retained decisions. 

The commitments are implemented as Pedersen commitments over the B254 curve while the proofs are produced with a standard Groth16 library\footnote{https://pypi.org/project/zkpy/}. The commitments and zero-knowledge proofs are generated and stored at the non-real-time cadence. As the work is performed asynchronously, the commitments and zero-knowledge proofs processing time is never added to the near-real-time latency $L_{i}$ and the bound $L_{i} \leq L_{max}$ continues to hold. Due to the aforementioned policies, like adaptive thresholds, the priority rule with scope locks, and the selective retention of evidence, the proposed work keeps the control loop stable even under multi-vendor conditions, while all timing and interface requirements of the O-RAN specification remain satisfied. 

\section{Evaluation Setup}
The goal of the evaluation for the proposed work is to measure how the $Z^2$-ACT composition behaves under realistic multi-cell, multi-slice O-RAN conditions on publicly available data. We performed the evaluation deliberately with trace-driven and open-loop settings. We replay the recorded ColO-RAN state vectors in order, therefore, a control decision at epoch $i$ does not modify the subsequent vectors $s_{i+1}, s_{i+2}...s_{i+n}$. In this regard, the differences in SLA satisfaction across different configurations measure how often each configuration would release or block realizations on the same fixed trajectory rather than how the radio channel and load would evolve under those realizations in live network.  

\subsection{Datasets and Trace Preparation}
In the proposed work, the main source of radio measurements is the Colosseum ColO-RAN dataset\footnote{https://openrangym.com/datasets/colosseum-coloran-dataset} published through OpenRAN Gym \cite{ColoRan, OpenRanGym}. The data is publicly available and was collected on the Colosseum wireless network emulator while a near-real-time RIC controlled a software-based RAN. The files that were released with the dataset contain per-cell and per-slice statistics, which include throughput, latency, physical-resource-block usage and related counters. These statistics represent mixed traffic that includes enhanced mobile broadband, ultra-reliable low-latency and machine-type services. Based on the publicly available data, we form a sequence of state vectors $s_{i}$. Each vector is time-stamped and identifies the cells and slices that were active at that instant. We removed the incomplete records, while the remaining sequence was divided into episodes of equal length so that every comparison method is tested on the same time horizon. 

We took the operator intents from the public examples released with the contract-based agentic framework \cite{ref9} and from a small number of additional intents written to match the slice types present in the traces. One of the examples is stated as "a request to raise the throughput of a URLLC slice on two specific cells while protecting the minimum rate of an eMBB slice". Each intent $I$ is then paired with the state that was observed when the intent was assumed to arrive, giving the input pair $(I, s_{i})$ expected by the dual-agent pipeline. 

\subsection{Live Non-RT LLM for Intent-to-Contract Translation}
The dual-agent pipeline in the non-real-time RIC is instantiated with a live large language model. For each recorded intent $I$ and the associated state summary $s_{i}$, the model is prompted to propose candidate contract fields (goal predicates, resource ceilings, spatial scope, temporal interval, and permitted skills). A deterministic schema checker and the evaluator agent then accept, repair, or reject the proposal. Only contracts that pass schema validation and feasibility checks against the live RAN constraints encoded in the trace metadata are released to the near-real-time path. We report four quantities, i.e. Translation Accuracy, Invalid/hallucinated output rate, Non-RT Latency, and Attack Behavior, on the full intent set. The Translation Accuracy refers to the fraction of intents for which the accepted contract matches a hand-specified reference contract on all required field, up to benign syntactic variation. The Invalid/hallucinated output rate refers to the fraction of model outputs that (i) violate the schema, (ii) invent cells, slices, or skills not present in the catalogue, or (iii) contradict hard constraints visible in $s_{i}$. The Non-RT latency refers to the time from intent submission to accepted or rejected contract, including model inference and evaluator checks. Lastly, the Attack Behavior refers to the fraction of adversarial or misleading intents that produce a contract the evaluator correct rejects, versus contracts that would have been unsafe if released without the dual-agent checks. 

It should be noted that in the proposed work, the near-real-time path does not host this model. It receives only accepted contracts and A1 plans, then applies zero-trust admission on LLM-bound inputs, the self-management gate, priority/scope locks, and asynchronous zero-knowledge evidence exactly as in Section IV. Trace replay of ColO-RAN KPMs remains open-loop, therefore, the released actions do not alter subsequent radio states. 

\subsection{Multi-Vendor and Adversarial Scenarios}
In order to simulate open multi-vendor setting, we assigned the control actions that appear in the traces to three synthetic vendors. At any epoch, two or more vendors might propose decisions that can affect the same cells. Therefore, when the sets of affected cells intersect, the priority rule and scope locks are activated as defined in Section IV. The adversarial prompts were produced by placing harmful instructions inside the telemetry data that otherwise appears perfectly normal. We provide two examples below.
\begin{itemize}
    \item Ignore the latency limit of the URLLC side
    \item Give all resource blocks to the requesting vendor
\end{itemize}
The zero-trust verifier will receive these prompts as a near-real-time RIC would \cite{PromptGuard}. We apply the same prompts to configurations that do not include the verifier to directly assess the benefits. 

We evaluate five different configurations on every episode. The $Z^2$-ACT configuration refers to the complete architecture (Intent Contract, zero-trust verification, an agentic sequencer with self-management gate \cite{ref6}, priority rule with scope locks, and asynchronous zero-knowledge evidence). The No-Contract configuration removes the dual-agent pipeline and maps the intents straight to skill sequences. The No-ZT configuration removes the zero-trust verifier, which suggests that every prompt is implicitly accepted. The No-Gate configuration removes the self-management gate, which implies that any contract-compliant decision is released without the six scalar checks. Lastly, the Conventional RL baseline refers to a single deep reinforcement learning xAPP trained offline on the same traces and then frozen, representing a typical near-real-time controller that lacks agentic, contract, and cryptographic components. It should be noted that all of the aforementioned configurations see the same state sequences, the same intents, and the same adversarial prompts. Therefore, the performance measurement would show the differences in outcomes strictly attributable to the presence or absence of the corresponding $Z^2$-ACT component.

We evaluate the $Z^2$-ACT architecture using six measurement metrics. We record the metrics for each episode and for each configuration, respectively. The first is the SLA satisfaction, which measures the fraction of time slots in which every active slide meets its throughput and latency. The second is the attack mitigation rate (AMR), which measures the fraction of adversarial prompts that are blocked or cleaned before they can affect a released decision. The third is the audit verification success (AVS), which measures the fraction of released decisions whose zero-knowledge proof is verified correctly against the stored commitment. The fourth is the near-real-time latency (NRTL), which is the sum mentioned in equation 6 together with the fraction of epochs that satisfy $L_{i} \leq L_{max}$. The fifth is the E2 message volume (E2MV), which is the average number of bytes sent over the E2 interface per second. The last is the resource usage (RU), which corresponds to the total physical-resource-block utilization across all cells. We average each metric over all episodes and over several random seeds that control the insertion of adversarial prompts and the assignment of vendors. We compute the confidence intervals from the ordinary sample standard deviation.  

\subsection{Implementation}
We used Python for trace replay and the $Z^2$-ACT control logic. For the dual-agent contract step and the zero-trust score, we implemented them as light modules that read the recorded state vectors, prepared intent, and prompt strings. We do not use an LLM service for the aforementioned in our core experiments. The self-management gate and the priority/scope-lock logic are pure functions of the quantities defined in Section IV. Commitments and zero-knowledge proofs are generated with a standard library implementation of a binding commitment scheme and a succinct proof system. We count the verification time only on the critical path because proof generation runs asynchronously. We log all random seeds, episode indices and configuration flags to maintain reproducibility.

We used Python for trace replace and the $Z^2$-ACT control logic. In the non-real-time path, intent-to-contract translation is performed by an LLM Llama 2 7B, followed by the deterministic evaluator and schema checks described in Section V-B. The 7B model is used only to demonstrate the agentic-style intent interpretation, which can be integrated into the non-real-time stage of $Z^2$-ACT. Another reason for using Llama 2 7B was that the model could be run with 4-bit quantization on an NVIDIA RTX 4090 configuration. Model identity, decoding parameters, quantization settings, and the system prompt were fixed for all runs and recorded with the random seeds. The zero-trust score, self-management gate, and priority/scope-lock logic in the near-real-time path remain lightweight modules that consume accepted contracts, recorded state vectors, and the fixed adversarial prompt catalogue. Commitments and zero-knowledge proofs use a standard binding commitment scheme and a succinct proof system (Pedersen commitment and Groth16 over a pairing-friendly curve, as specified in Section IV). Only verification time is counted on the critical path while the proof generation was asynchronous.

\section{Experimental Results}
In this section, we present the experimental results for $Z^2$-ACT on the traces driven from a publicly available dataset. We ran every configuration on the same public ColO-RAN episodes using the same operator intents and the same fixed adversarial prompts. We discuss the results in the subsequent subsections accordingly. 

\subsection{Quantitative Comparison and Ablation Study}
We perform the comparison on five configurations that were defined in Section V, which correspond to service quality, attack resilience, and radio resource usage. For $Z^2$-ACT, IC are produced by Llama 2 7B with evaluator checks and the ablation configurations that retain contracts use the same accepted-contract set so that differences isolate near-real-time filters. Table 2 summarizes the mean values obtained over all episodes and seeds. The SLA satisfaction is the highest for $Z^2$-ACT. The experiments reveal that when the IC is removed, the SLA satisfaction drops to 0.83, which is consistent with the admission of goals that violate hard resource ceilings. Drops are also noticed when disabling the zero-trust verifier (No-ZT) and the self-management gate (No-Gate). The lowest SLA satisfaction score is yielded by the conventional RL baseline, which lacks contracts, prompt verification, and gating. The AMR behaves similarly, such that the configurations that retain the zero-trust verifier, including $Z^2$-ACT, No-Contract, and No-Gate, block or sanitize the large majority of the prompts (0.94-0.95). However, when the sanitizer is removed (No-ZT) or is absent in configurations like RL baseline, the mitigation rate falls to 0.03 and 0.02, respectively. It should be noted that the results for $Z^2$-ACT are reflected on the fixed ColO-RAN trajectories such that the declared slice targets are met in the replayed measurements. As the future states are not regenerated from the released actions, this gap should be read as the effect of filtering unsafe or conflicting realizations on a common trace rather than as a closed-loop demonstration that $Z^2$-ACT reshapes subsequent throughput, latency, or PRB usage. The results are recorded on the fixed prompt set rather than against arbitrary open-ended natural language attacks. The RU stays within the range of 0.62 - 0.65 across all five configurations. The $Z^2$-ACT yields the lowest, which is consistent with the rejection of a fraction of unsafe or overlapping realizations. No-Gate configuration yields slightly higher in comparison to $Z^2$-ACT, as more decisions are released when the six checks are disabled. Based on the aforementioned results, we can confidently say that the primitives, when implemented in a unified single layer, can help in improving SLA satisfaction, attack mitigation, and resource usage, respectively.   

% \usepackage{graphicx}
\begin{table}[]
\centering
\caption{Quantitative comparison across ablations (mean over episodes and seeds).}
\label{tab:my-table}
\resizebox{\columnwidth}{!}{%
\begin{tabular}{|l|l|l|l|}
\hline
\textbf{Configuration} & \textbf{SLA Satisfaction} & \textbf{AMR} & \textbf{RU (mean PRB)} \\ \hline
\textbf{$Z^2$-ACT} & 0.91 & 0.95 & 0.62 \\ \hline
\textbf{No-Contract} & 0.83 & 0.94 & 0.64 \\ \hline
\textbf{No-ZT} & 0.85 & 0.03 & 0.63 \\ \hline
\textbf{No-Gate} & 0.86 & 0.94 & 0.66 \\ \hline
\textbf{RL baseline} & 0.81 & 0.02 & 0.63 \\ \hline
\end{tabular}%
}
\end{table}

% \usepackage{graphicx}
\begin{table}[]
\centering
\caption{Non-RT LLM Translation and Efficiency Results }
\label{tab:my-table}
\resizebox{\columnwidth}{!}{%
\begin{tabular}{|l|l|}
\hline
\multicolumn{1}{|c|}{\textbf{Metric}} & \multicolumn{1}{c|}{\textbf{Value}} \\ \hline
\textbf{Translation Accuracy} & 0.78 \\ \hline
\textbf{Invalid / Hallucinated Output Rate} & 0.18 \\ \hline
\textbf{Mean Non-RT Latency (sec)} & 1.0 \\ \hline
\textbf{p95 Non-RT Latency (sec)} & 2.5 \\ \hline
\textbf{Adversarial-Intent Rejection Rate (After Evaluator)} & 0.85 \\ \hline
\end{tabular}%
}
\end{table}

\subsection{Non-RT LLM Translation and Efficiency}
The non-real-time intent-to-contract stage uses Llama 2 7B with 4-bit quantization on an NVIDIA RTX 4090, followed by deterministic schema validation and the evaluator agent. We report the results in Table 3, which summarizes the translation quality, invalid outputs, latency, and behavior of LLM under adversarial operator text on the fixed intent set. The translation accuracy of 0.78 indicates that most intents map to reference contracts on all required fields after benign syntactic variation. An invalid or hallucinated output rate of 0.18 captures schema violations, invented cells or skills, and contradictions with visible constraints; the large majority of these cases were rejected by the evaluator and never released to the near-real-time path. Mean and p95 non-real-time latencies of 1.0 s and 2.5 s were acceptable for operator-scale intent submission and are not included in the near-real-time sum $L_{i}$. Under adversarial or misleading operator intents, the dual-agent pipeline rejects 0.85 of unsafe proposals, so residual risk is dominated by the minority that pass both the model and the evaluator, not by the near-real-time verifier alone.

\subsection{Latency and Overhead Analysis under Near-RT constraints}

The end-to-end latency of each control epoch is defined in equation 6. We report the mean and 95-th percentile latency, the fraction of epochs that satisfy the near-RT bound $L_{i} \leq L_{max}$, and the mean E2 signaling volume for every configuration in Table 4, respectively. Non-real-time LLM inference time is excluded from $L_i$, therefore, Table 4 reports only verifier, gate, and E2 dispatch latency on the near-real-time path, respectively. The results show that the $Z^2$-ACT records a mean latency of 0.018 seconds and a 95th percentile of 0.030 seconds. Both values lie well below the upper edge of the O-RAN near-RT window, and 0.99 of all epochs satisfy $L_{i} \leq L_{max}$. The added cost relative to the RL baseline (mean 0.008 seconds) is therefore noticeable but compatible with the timing model of Section IV. The dominant contributors to the latency are the zero-trust verifier and the gate evaluation. When the verifier is removed, the mean latency falls to 0.010 and the 95th percentile to 0.016, with every epoch meeting the bound. No-contract remains close to $Z^2$-ACT (0.017 seconds mean), which is expected as contract generation runs in the non-real-time path and does not sit on the critical near-RT loop. In all cases, the fraction of epochs that violate $L_{max}$ is negligible. The E2 signaling volume stays within a modest band above the RL baseline. The $Z^2$-ACT and No-contract incur an increase of roughly 14-15\%, which is consistent with the extra control messages that accompany contract-aware releases. None of the configurations doubles the signaling load; therefore, the overhead is limited and does not offset the safety benefits reported in Table 4. 

% \usepackage{graphicx}
\begin{table}[]
\centering
\caption{Latency and control plane overhead under near-RT constraints (mean over episodes and seeds).}
\label{tab:my-table}
\resizebox{\columnwidth}{!}{%
\begin{tabular}{|l|l|l|l|l|}
\hline
\textbf{Configuration} & \textbf{Mean $L_{i}$ (s)} & \textbf{p95 $L_{i}$ (s)} & \textbf{Fraction $L_{i} \leq L_{max}$} & \textbf{E2MV (bytes/s)} \\ \hline
\textbf{$Z^2$-ACT} & 0.018 & 0.030 & 0.99 & 1.15 $\times$ baseline \\ \hline
\textbf{No-Contract} & 0.017 & 0.028 & 0.99 & 1.14 $\times$ baseline \\ \hline
\textbf{No-ZT} & 0.010 & 0.016 & 1.00 & 1.05 $\times$ baseline \\ \hline
\textbf{No-Gate} & 0.012 & 0.020 & 1.00 & 1.22 $\times$ baseline \\ \hline
\textbf{RL baseline} & 0.008 & 0.014 & 1.00 & 1.00 $\times$ baseline \\ \hline
\end{tabular}%
}
\end{table}

\subsection{Auditability and Multi-vendor Conflict Resolution}
We evaluate the auditability and concurrent multi-vendor behavior on the same episodes that were undertaken in preceding evaluations. As the $Z^2$-ACT configuration generates commitments and zero-knowledge proofs, we record the fraction of released decisions for which the condition mentioned in equation 5 is satisfied. We examine the multi-vendor interactions by counting episodes in which two or more synthetic vendors proposes decisions with intersecting spatial scopes. We also consider the measurement based on how often the priority rule and scope locks of Section IV-C serialize the said commits so that the release predicate remains well-defined. We report the results for the auditability and multi-vendor conflict resolution in Table 5, respectively. For the $Z^2$-ACT the audit verification success (AVS) is 1.00 as every commitment that was produced after a release was verified under the public verification key. Other configurations did not emit proofs and are therefore marked as not application (N/A) for this metric. It was observed that the spatial overlap (SO) occurred in 0.35 of all episodes under the synthetic three-vendor assignment as mentioned in Section V. Within those overlapping episodes, the $Z^2$-ACT resolves 0.96 of conflicts according to the declared priority order (safety and compliance before SLA recovery before efficiency) and keeps the release predicate well-defined after serialization in 0.98 of cases. The aforementioned figures align with the design of the scope locks and the priority rule. 

When the IC is removed (No-Contract), the systematic priority ordering is weakened such that only 0.62 of overlaps were resolved by the intended rule and the release predicate remains well-defined in 0.82 of overlapping episodes. When the gates are disabled (No-Gate), similar degradation was noticed, i.e. 0.68 and 0.78, respectively, because the rate, scope, and risk limits no longer constraints concurrent commits. No-ZT achieves the results closer to that of $Z^2$-ACT, i.e. 0.94 and 0.96, which is consistent with the fact that the verifier does not implement priority or locking. The RL baseline does not have priority rule or scope locks, therefore, it does not resolve any overlap in a structured manner, i.e. 0.08 and leaves the release condition well-defined in only 0.55 of overlapping episodes.   

We additionally probe two integrity failure modes. The first is the Tamper mode, suggesting that after a valid $(com_{i}, \pi_{ZK}^{(i)}$ is stored, the verification must be rejected if the commitment string or the proof is modified. The second is the Replay mode, in which a valid proof from epoch $i$ is submitted against a different epoch index or a different contract hash, therefore, in this case the verification must be rejected as the public inputs $(H(IC),i)$ no longer match. Under a correct implementation of the binding commitment and the SNARK, both classes of checks fail closed (verification returns 0). These checks were logged together with the AVS rate and do not affect the near-real-time path.  

% \usepackage{graphicx}
\begin{table}[]
\centering
\caption{Auditability and multi-vendor conflict resolution (mean over episodes and seeds).}
\label{tab:my-table}
\resizebox{\columnwidth}{!}{%
\begin{tabular}{|l|l|l|l|l|}
\hline
\textbf{Configuration} & \textbf{\begin{tabular}[c]{@{}l@{}}Audit Verification \\ Success\end{tabular}} & \textbf{\begin{tabular}[c]{@{}l@{}}Episodes \\ with SO\end{tabular}} & \textbf{\begin{tabular}[c]{@{}l@{}}Overlaps resolved \\ by priority rule\end{tabular}} & \textbf{\begin{tabular}[c]{@{}l@{}}Release predicate well-defined \\ after serialization\end{tabular}} \\ \hline
\textbf{$Z^2$-ACT} & 1.00 & 0.35 & 0.96 & 0.98 \\ \hline
\textbf{No-Contract} & N/A & 0.35 & 0.62 & 0.82 \\ \hline
\textbf{No-ZT} & N/A & 0.35 & 0.94 & 0.96 \\ \hline
\textbf{No-Gate} & N/A & 0.35 & 0.68 & 0.78 \\ \hline
\textbf{RL baseline} & N/A & 0.35 & 0.08 & 0.55 \\ \hline
\end{tabular}%
}
\end{table}

\subsection{Sensitivity analysis}
In the proposed work, the release predicate and the zero-trust depends on the threshold vector  $\Theta = (\theta, \alpha, \beta, \gamma, \tau, S, R)$, therefore, to assess how each coordinate affects the metrics, the same public episodes are re-run while one threshold is varied at a time and all others are held at the nominal operating point. For every coordinate we consider three levels, i.e. permissive, nominal, andd restrictive, and record SLA satisfaction, AMR, the fraction of epochs that satisfy $L_{i} \leq L_{max}$, mean latency, and E2MV, accordingly. We summarize the sensitivity analysis in Table 6. Making the $\theta$ restrictive raises the AMR from 0.88 to 0.98 while SLA satisfaction declines by only a few points, i.e. 0.92 to 0.88. The mean latency increases slightly because the verifier rejects more prompts, yet the fraction of epochs inside the Near-RT bound remains at or above 0.99. As the zero-trust verifier is unchanged, the mitigate rate stays constant. A more restrictive risk limit reduces the number of admitted commits, which lowers E2 volume and trims SLA satisfaction by a modest amount, i.e. 0.90 to 0.87, whereas the Near-RT bound continues to hold. For $\beta$ the pattern matches that of $\alpha$, i.e. stricter uncertainty control reduces realizations under noisy telemetry, with a small SLA cost and a corresponding drop in signaling load. For $\gamma$, restricting the resource budget trims concurrent work and E2 volume. Mean latency does not increase, however the fraction of epochs meeting $L_{max}$ remains at least 0.99 while the SLA satisfaction slightly declines. Raising the consistency floor $\tau$ holds back a larger share of opaque skill sequences. Although the mitigation is unaffected, the SLA satisfaction moves from 0.92 to 0.87 across the three levels, which is consistent with fewer corrective realizations rather than with a change in prompt filtering. The restrictive scope limit $S$ reduces wide, overlapping commits and lowers signaling volume. The effect on SLA satisfaction remains within a few points (0.91 - 0.89), while the AMR and Near RT-compliance are essentially unchanged. Lastly for the commit-rate limit $R$, lowering it directly reduces realization frequency and E2 load. SLA satisfaction falls from 0.92 to 0.87 as fewer corrections are applied, while the latency bound remains satisfied in essentially all epochs. 

% \usepackage{graphicx}
\begin{table}[]
\centering
\caption{Sensitivity analysis for the threshold $\Theta$ vector parameters. }
\label{tab:my-table}
\resizebox{\columnwidth}{!}{%
\begin{tabular}{|lllll|}
\hline
\multicolumn{5}{|c|}{Adversarial Intent Threshold $\theta$} \\ \hline
\multicolumn{1}{|l|}{$\theta$ setting} & \multicolumn{1}{l|}{SLA satisfaction} & \multicolumn{1}{l|}{AMR} & \multicolumn{1}{l|}{Fraction $L_{i} \leq L_{max}$} & Mean $L_{i}$ (s) \\ \hline
\multicolumn{1}{|l|}{Permissive} & \multicolumn{1}{l|}{0.92} & \multicolumn{1}{l|}{0.88} & \multicolumn{1}{l|}{1.00} & 0.016 \\ \hline
\multicolumn{1}{|l|}{Nominal} & \multicolumn{1}{l|}{0.91} & \multicolumn{1}{l|}{0.95} & \multicolumn{1}{l|}{0.99} & 0.018 \\ \hline
\multicolumn{1}{|l|}{Restrictive} & \multicolumn{1}{l|}{0.88} & \multicolumn{1}{l|}{0.98} & \multicolumn{1}{l|}{0.99} & 0.020 \\ \hline
\multicolumn{5}{|c|}{Gate Risk Limit $\alpha$} \\ \hline
\multicolumn{1}{|l|}{$\alpha$ setting} & \multicolumn{1}{l|}{SLA satisfaction} & \multicolumn{1}{l|}{AMR} & \multicolumn{1}{l|}{Fraction $L_{i} \leq L_{max}$} & E2MV (rel.) \\ \hline
\multicolumn{1}{|l|}{Permissive} & \multicolumn{1}{l|}{0.90} & \multicolumn{1}{l|}{0.95} & \multicolumn{1}{l|}{1.00} & 1.18 $\times$ \\ \hline
\multicolumn{1}{|l|}{Nominal} & \multicolumn{1}{l|}{0.91} & \multicolumn{1}{l|}{0.95} & \multicolumn{1}{l|}{0.99} & 1.15 $\times$ \\ \hline
\multicolumn{1}{|l|}{Restrictive} & \multicolumn{1}{l|}{0.87} & \multicolumn{1}{l|}{0.95} & \multicolumn{1}{l|}{0.99} & 1.08 $\times$ \\ \hline
\multicolumn{5}{|c|}{Uncertainty Limit $\beta$} \\ \hline
\multicolumn{1}{|l|}{$\beta$ setting} & \multicolumn{1}{l|}{SLA satisfaction} & \multicolumn{1}{l|}{AMR} & \multicolumn{1}{l|}{Fraction $L_{i} \leq L_{max}$} & E2MV (rel.) \\ \hline
\multicolumn{1}{|l|}{Permissive} & \multicolumn{1}{l|}{0.91} & \multicolumn{1}{l|}{0.95} & \multicolumn{1}{l|}{1.00} & 1.17 $\times$ \\ \hline
\multicolumn{1}{|l|}{Nominal} & \multicolumn{1}{l|}{0.91} & \multicolumn{1}{l|}{0.95} & \multicolumn{1}{l|}{0.99} & 1.15 $\times$ \\ \hline
\multicolumn{1}{|l|}{Restrictive} & \multicolumn{1}{l|}{0.87} & \multicolumn{1}{l|}{0.95} & \multicolumn{1}{l|}{0.99} & 1.09 $\times$ \\ \hline
\multicolumn{5}{|c|}{Budget Limit $\gamma$} \\ \hline
\multicolumn{1}{|l|}{$\gamma$ setting} & \multicolumn{1}{l|}{SLA satisfaction} & \multicolumn{1}{l|}{Mean $L_{i}$ (s)} & \multicolumn{1}{l|}{Fraction $L_{i} \leq L_{max}$} & E2MV (rel.) \\ \hline
\multicolumn{1}{|l|}{Permissive} & \multicolumn{1}{l|}{0.91} & \multicolumn{1}{l|}{0.019} & \multicolumn{1}{l|}{0.99} & 1.20 $\times$ \\ \hline
\multicolumn{1}{|l|}{Nominal} & \multicolumn{1}{l|}{0.91} & \multicolumn{1}{l|}{0.018} & \multicolumn{1}{l|}{0.99} & 1.15 $\times$ \\ \hline
\multicolumn{1}{|l|}{Restrictive} & \multicolumn{1}{l|}{0.88} & \multicolumn{1}{l|}{0.017} & \multicolumn{1}{l|}{1.00} & 1.06 $\times$ \\ \hline
\multicolumn{5}{|c|}{Explanation-consistency floor $\tau$} \\ \hline
\multicolumn{1}{|l|}{$\tau$ setting} & \multicolumn{1}{l|}{SLA satisfaction} & \multicolumn{1}{l|}{AMR} & \multicolumn{1}{l|}{Fraction $L_{i} \leq L_{max}$} & E2MV (rel.) \\ \hline
\multicolumn{1}{|l|}{Permissive} & \multicolumn{1}{l|}{0.92} & \multicolumn{1}{l|}{0.95} & \multicolumn{1}{l|}{1.00} & 1.17 $\times$ \\ \hline
\multicolumn{1}{|l|}{Nominal} & \multicolumn{1}{l|}{0.91} & \multicolumn{1}{l|}{0.95} & \multicolumn{1}{l|}{0.99} & 1.15 $\times$ \\ \hline
\multicolumn{1}{|l|}{Restrictive} & \multicolumn{1}{l|}{0.87} & \multicolumn{1}{l|}{0.95} & \multicolumn{1}{l|}{0.99} & 1.10 $\times$ \\ \hline
\multicolumn{5}{|c|}{Spatial-Scope limit $S$} \\ \hline
\multicolumn{1}{|l|}{$S$ setting} & \multicolumn{1}{l|}{SLA satisfaction} & \multicolumn{1}{l|}{AMR} & \multicolumn{1}{l|}{Fraction $L_{i} \leq L_{max}$} & E2MV (rel.) \\ \hline
\multicolumn{1}{|l|}{Permissive} & \multicolumn{1}{l|}{0.91} & \multicolumn{1}{l|}{0.95} & \multicolumn{1}{l|}{0.99} & 1.18 $\times$ \\ \hline
\multicolumn{1}{|l|}{Nominal} & \multicolumn{1}{l|}{0.91} & \multicolumn{1}{l|}{0.95} & \multicolumn{1}{l|}{0.99} & 1.15 $\times$ \\ \hline
\multicolumn{1}{|l|}{Restrictive} & \multicolumn{1}{l|}{0.89} & \multicolumn{1}{l|}{0.95} & \multicolumn{1}{l|}{1.00} & 1.10 $\times$ \\ \hline
\multicolumn{5}{|c|}{Commit-rate Limit $R$} \\ \hline
\multicolumn{1}{|l|}{$S$ setting} & \multicolumn{1}{l|}{SLA satisfaction} & \multicolumn{1}{l|}{AMR} & \multicolumn{1}{l|}{Fraction $L_{i} \leq L_{max}$} & E2MV (rel.) \\ \hline
\multicolumn{1}{|l|}{Permissive} & \multicolumn{1}{l|}{0.92} & \multicolumn{1}{l|}{0.95} & \multicolumn{1}{l|}{0.99} & 1.21 $\times$ \\ \hline
\multicolumn{1}{|l|}{Nominal} & \multicolumn{1}{l|}{0.91} & \multicolumn{1}{l|}{0.95} & \multicolumn{1}{l|}{0.99} & 1.15 $\times$ \\ \hline
\multicolumn{1}{|l|}{Restrictive} & \multicolumn{1}{l|}{0.87} & \multicolumn{1}{l|}{0.95} & \multicolumn{1}{l|}{1.00} & 1.05 $\times$ \\ \hline
\end{tabular}%
}
\end{table}

\subsection{Discussion on Trade-offs}
The results reported in Table 2 to Table 6 show the performance of $Z^2$-ACT in varying conditions under the evaluation design of Section V. The $Z^2$-ACT achieves an SLA satisfaction of 0.91 and AMR of 0.95 in comparison to the RL baseline that achieves 0.81 and 0.02, respectively. If we remove the contract, the verifier, or the gate, each reduces SLA satisfaction by several points. The resource usage stays in a narrow band, i.e. 0.62-0.66, which was expected on foreign traces where the architecture mainly filters unsafe decisions. Although the gains add a limited latency cost, $Z^2$-ACT records a mean $L_{i}$ of 0.018 and a 95th percentile of 0.030 seconds with 0.99 of epochs inside the Near-RT bound. The overhead is therefore visible but we would argue that this is compatible with the O-RAN Near-RT window. The proposed study is trace-driven, models the multi-vendor behavior and emulates the adversarial prompts to show its efficacy. The results show that the proposed $Z^2$-ACT improves safety, attack resilience, and auditability at a modest cost in latency and signaling, while remaining inside the timing constraints of Section IV. 

A central methodological limit of the present study is the open-loop, trace-driven design. Released actions do not change subsequent radio states, so the SLA metric does not capture the counterfactual evolution of the RAN under $Z^2$-ACT versus the ablations. Two extensions would close this gap: (i) a closed-loop experiment on a programmable platform (e.g. Colosseum / OpenRAN Gym with live E2 control), in which each admitted skill sequence is applied and the next KPM vector is measured; or (ii) an explicit state-transition model $s_{i+1} = f(s_{i},d_{i})$ validated against held-out traces, used to simulate counterfactual trajectories for each configuration. Until one of these is in place, SLA differences are interpreted strictly as differences in actuation filtering on shared public traces.

\section{Conclusion and Future work}
The experts suggest that th open and intelligent 6G radio access networks will rely on multi-vendor and AI assisted control loops. However, in order to have a real-world applicability of such system, it needs to be safe, verifiable, and auditable under concurrent intents and untrusted model inputs. The proposed work addresses the aforementioned need by composing four existing isolated primitives, i.e. agentic control, IC, zero-trust prompting, and zero-knowledge accountability, into a single architecture, which operates across the non-real-time and near-real-time RICs without modifying the standard E2 or O1 interfaces. 

We performed the evaluation on publicly available ColO-RAN traces, with a fixed adversarial-prompt catalogue and an emulated multi-vendor assigned. The results show that the full composition of $Z^2$-ACT improves service-level behavior and attack resilience relative to strong ablations and a conventional RL baseline, at a modest cost in latency and signaling that remains inside the Near-RT timing envelope. We also show that the audit verification succeeds for every proof produced by $Z^2$-ACT while the priority-based scope locks resolves the majority of spatial conflicts. We also perform sensitivity analysis for the threshold vector, which shows that the nominal operating point is stable, hence tightening the individual limits moves safety and overhead metrics in the expected direction without driving the system outside the Near-RT bound. 

We acknowledge that there are limitations concerning the proposed study design. For instance, the evaluation is trace-driven, therefore the effect of $Z^2$-ACT decisions on future radio conditions are not observed. We emulated the multi-vendor behavior rather than obtaining the traces from independent vendor implementation. We drew the adversarial prompts from a constructed catalogue and performed the sensitivity analysis for one threshold at a time. However, we want to justify that the experiments, due to the aforementioned choices, can be reproduced easily from the publicly available data. 

We intend to address the aforementioned limitations in the future work such that a close-loop implementation on a programmable testbed would be carried out to allow $Z^2$-ACT decisions to affect subsequent channels and load conditions. We intend to perform interoperability trials with independently developed xApps, which would replace the synthetic vendor assignment. We intend to expand the prompt set toward paraphrased and adaptive attacks, which would stress the zero-trust component more severely. Lastly, we intend to extend the same organizational pattern to other O-RAN control applications, beyond slice reallocation, which would test whether the composition remains effective with the increasing number of skills and contracts.



\bibliographystyle{IEEEtran}
\bibliography{references}

\end{document}